%% file: main.tex
\documentclass{article}

\usepackage{arxiv}

\usepackage[utf8]{inputenc} % allow utf-8 input
\usepackage[T1]{fontenc}    % use 8-bit T1 fonts
\usepackage{hyperref}       % hyperlinks
\usepackage{url}            % simple URL typesetting
\usepackage{booktabs}       % professional-quality tables
\usepackage{amsfonts}       % blackboard math symbols
\usepackage{nicefrac}       % compact symbols for 1/2, etc.
\usepackage{microtype}      % microtypography
\usepackage{lipsum}
\usepackage{graphicx}
\usepackage{amsmath}
\usepackage{multirow}
\usepackage{makecell}
\usepackage{xspace}
\usepackage{subcaption}

\title{Uni-Mol Docking V2: Towards Realistic and Accurate Binding Pose Prediction}

\author{ DP Technology Technical Staff \\
        Beijing, Haidian District \\
	\texttt{dplc@dp.tech} \\
	\AND
	AI for Science Institute \\
        Beijing, Haidian District \\
}

\date{}

\renewcommand{\headeright}{}
\renewcommand{\undertitle}{}
\renewcommand{\shorttitle}{}

\begin{document}
\maketitle

\vspace{-20pt}
\begin{center}
\href{https://bohrium.dp.tech/apps/unimoldockingv2}{https://bohrium.dp.tech/apps/unimoldockingv2}
\end{center}

\vspace{20pt}

% keywords can be removed
% \keywords{Uni-Mol; Molecular Docking; Pretraining}

\input{sections/abstract}
\input{sections/intro}
\input{sections/method_and_result}
\input{sections/conclusion}
\input{sections/code_data}
\input{sections/contribution}

% \newpage
% {
% \small
% \printbibliography
% }
\bibliographystyle{unsrt}  
\bibliography{references}
\end{document}

%% file: sections/abstract.tex
\begin{abstract}
In recent years, machine learning (ML) methods have emerged as promising alternatives for molecular docking, offering the potential for high accuracy without incurring prohibitive computational costs. However, recent studies have indicated that these ML models may overfit to quantitative metrics while neglecting the physical constraints inherent in the problem. In this work, we present Uni-Mol Docking V2, which demonstrates a remarkable improvement in performance, accurately predicting the binding poses of 77+\% of ligands in the PoseBusters benchmark with an RMSD value of less than 2.0 \AA, and 75+\% passing all quality checks. This represents a significant increase from the 62\% achieved by the previous Uni-Mol Docking model. Notably, our Uni-Mol Docking approach generates chemically accurate predictions, circumventing issues such as chirality inversions and steric clashes that have plagued previous ML models. Furthermore, we observe enhanced performance in terms of high-quality predictions (RMSD values of less than 1.0 \AA~ and 1.5 \AA) and physical soundness when Uni-Mol Docking is combined with more physics-based methods like Uni-Dock. Our results represent a significant advancement in the application of artificial intelligence for scientific research, adopting a holistic approach to ligand docking that is well-suited for industrial applications in virtual screening and drug design. The code, data and service for Uni-Mol Docking are publicly available for use and further development in \hyperlink{Uni-Mol}{https://github.com/dptech-corp/Uni-Mol}.
\end{abstract}

%% file: sections/intro.tex
\section{Introduction}
The Uni-Mol modelling series \cite{zhou2023unimol} described the pretraining of general molecular encoders and showcased their applications in various 2D and 3D downstream tasks such as molecular conformation generation, molecular property prediction and molecular docking. The Uni-Mol paradigm was later extended to quantum property prediction in \cite{lu2023highly}, showcasing the quality and applicability of molecular features learned by the architecture. 

In molecular docking, leveraging a pretrained molecular encoder, pretrained pocket encoder and joint pocket-ligand blocks, UniMol Docking achieved superior performance when compared to traditional docking algorithms like Autodock Vina\cite{trott2010autodock} in the CASF-2016 benchmark \cite{su2018comparative}.

Recent work highlighted the need for critical evaluation of physical and chemical plausibility of docked posed by ML models, and showed that despite exhibiting better quantitative metrics such as \% of <2.0 \AA~RMSD, deep learning models did not perform significantly better than traditional docking programs. In that work, UniMol was reported to achieve a 22\% <2.0 \AA~RMSD on the PoseBusters set \cite{buttenschoen2024posebusters}. However, we attribute this low performance to suboptimal data processing and propose a standard pipeline in this work.

Lately, several works have proposed different approaches to ML docking shortcommings. On the model side, RFAA \cite{Krishna2023.10.09.561603} proposed an all-atom modelling and extended protein folding to proteins, nucleic acids, ions and ligands; DiffDock-pocket \cite{anonymous2023diffdockpocket} and Umol \cite{bryant2023structure} proposed to lower the need for crystal structures by including protein flxibility in the modelling; the latest AlphaFold \cite{alphalatest} report followed a similar approach and achieved the highest result on the PoseBusters benchmark, while also showcasing a partial improvement in chemical accuracy (stereochmistry). On the plausibility of results, light postprocessing as an artifact mitigation strategy \cite{alcaide2023umdfit} has been briefly explored, but it did not achieved a complete removal of unplausible results.

Therefore, this work introduces three main results: 
\begin{itemize}
\item A reproducible setup for molecular docking for the previously released Uni-Mol Docking, with correct dataset processing and reproducible results, achieving state of the art among publicly available deep-learning based molecular docking in CASF-2016, PoseBusters test set and Astex Diverse Set. Code is made publicly available (See Data \& Availability). 
\item Results from UniMol Docking V2 showcase increased performance on unseen data and achieved the best result to our knowledge on the PoseBusters test set as of November 22, 2023.
\item A step change in chemical accuracy of deep learning models, where all the unphysical and chemical issues presented previously for ML models' predictions have been corrected.
\end{itemize}

%% file: sections/method_and_result.tex
\section{Methodlogy}
\begin{figure*}[t]
    \centering
    \includegraphics[width=6.5in]{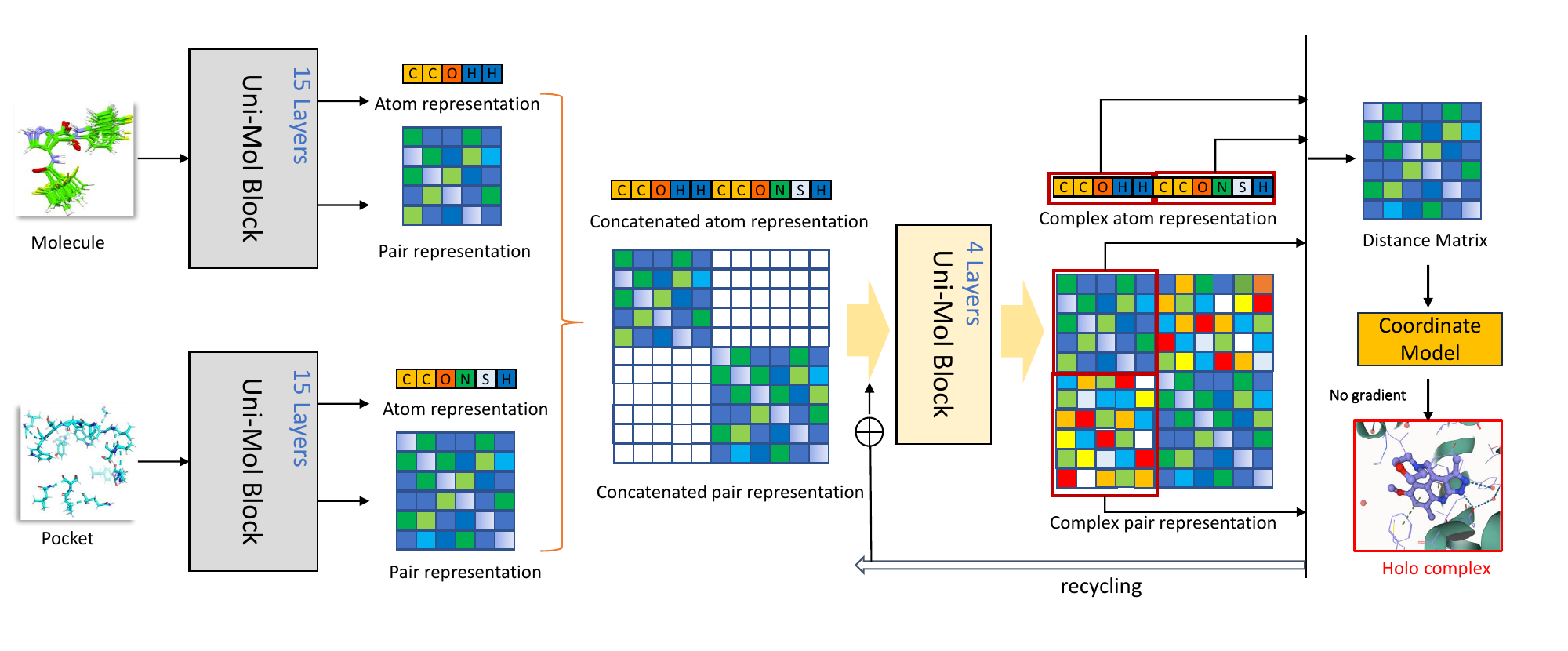}
    \caption{Framework of Uni-Mol Docking V2}
    \label{fig:Uni-Mol docking v2 architecture}
\end{figure*}

We collect protein-ligand binding data from MOAD~\cite{hu2005binding} for training. The protein data is prepared using a specific pipeline that includes the proper addition of correct hydrogen atoms, protonation information, and completion of missing heavy atoms and residues. We split the data into training and validation sets in a 9:1 ratio randomly. The training of Uni-Mol Docking V2 starts from the pretrained molecular and pocket checkpoints of Uni-Mol, the same as V1. We train the model for 100 epochs on 8 V100 GPUs with a batch size of 64, doubling the size compared to V1.  

The UniMol Docking V2 needs the same input as the previous version: a known pocket and the chemical compound to be docked. Based on this, the pocket is taken in cubic format of the ligand size and a margin of 10 \AA (similar to existing tools such as AutoDock Vina)  and a ligand conformer (which can be provided or automatically built from the ligand smiles using standard cheminformatics tools). The output is a 3D pose of the ligand bound to the protein of interest. By pre-computing pocket features UniMol Docking can be applied to virtual screening scenarios efficiently.

The combination of Uni-Mol Docking and UniDock is tailored for industrial applications and rational drug design usecases, where the binding pocket better characterized. UniDock further allows leveraging information from cofactors and crystalographic waters to further improve accuracy. This setup has been described in a recent preprint\cite{yang_lin_yuan_li_zou_zhou_zhang_zheng_2023} As claimed previously, the PoseBusters test set is composed of unseen data for UniMol (both previous and latest V2 version) as that only encompasses selected protein-ligand complexes released until 2019 included, whereas the test set is composed by structures released from 2020 onwards. We report results on both the PoseBusters benchmark set and Astex Diverse benchmark set, as introduced by \cite{buttenschoen2023posebusters}. 

% \subsection{Results}
\section{Results}
\subsection{Standard protocol for the previous version of Uni-Mol Docking}

After following a standardised protocol for inference of the PoseBusters and Astex test sets, the accuracy is substantially higher than that originally reported in the first versions of the preprint in \cite{buttenschoen2023posebusters}. The \% compounds predicted < 2.0 \AA RMSD of the ground truth is above 62\%. To our knowledge, this constitutes the best result on the PoseBusters set by an open source model by November 22, 2023.  
After discussion with \cite{buttenschoen2023posebusters} authors, we attribute the performance delta to differences in the data ingestion and input generation (the pocket is all atoms <6.0 \AA of any ligand heavy atom for this result, but was 8.0 \AA in early versions of \cite{buttenschoen2023posebusters}). The deep learning model architecture and weights are kept the same. To avoid such issues in the future, we make our code available, including a containerized environment, data files and scripts to assemble them, as well as instructions already available on model loading and inference.
Although plausibility has not been considered in this result, unphysical artifacts can be mitigated by following strategies such as \cite{alcaide2023umdfit}.

\subsection{Uni-Mol Docking V2 results}

We present the results obtained by the version of Uni-Mol Docking V2, achieving a staggering 77+\% of ligands in the PoseBusters benchmark predicted with < 2.0 \AA RMSD, and 75+\% of complexes passing all PoseBusters quality checks. This represents a new state of the art for ML-assisted protein-ligand docking.
\begin{figure*}[t]
    \centering
    \includegraphics[width=6.5in]{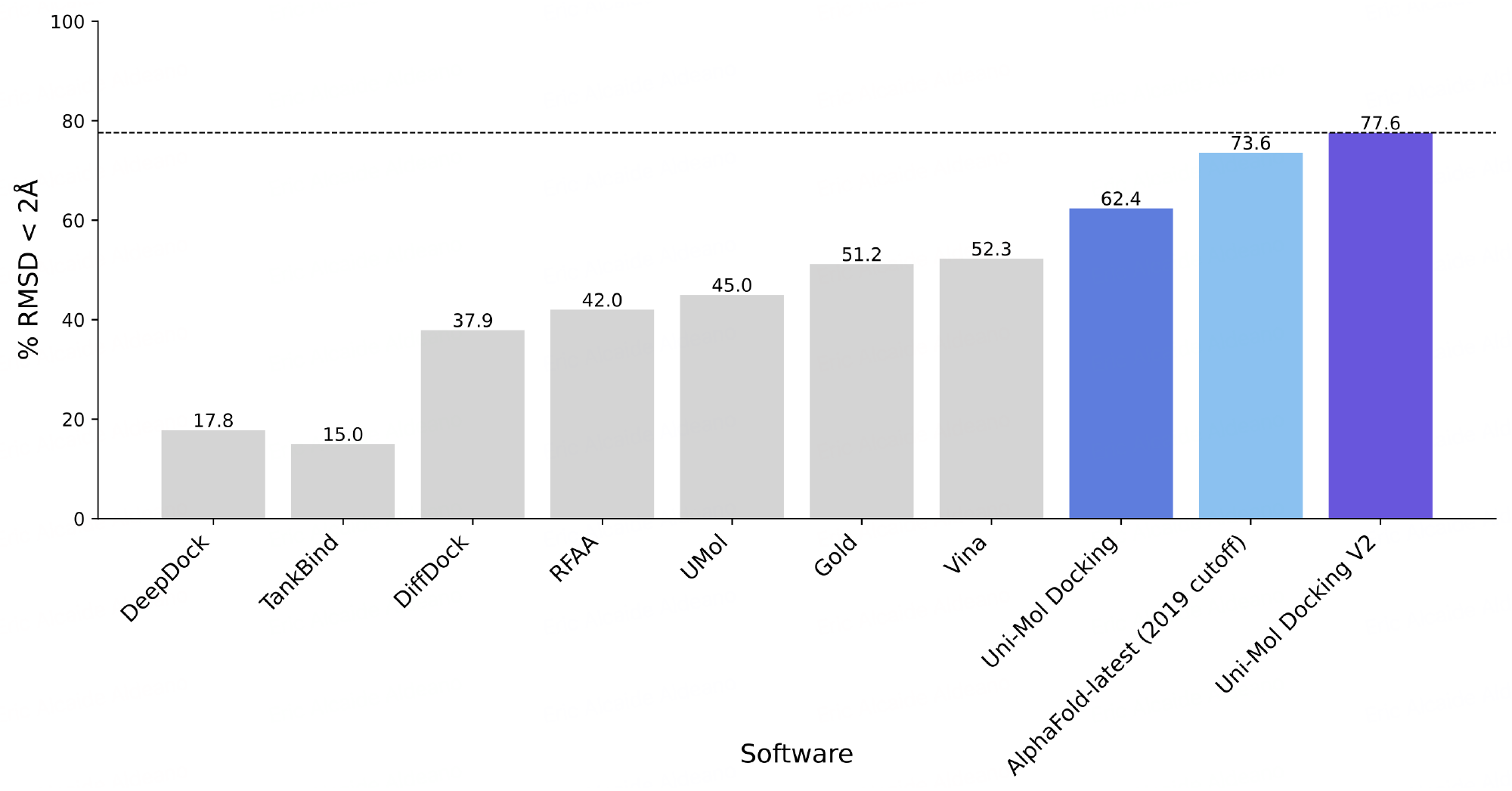}
    \caption{Performance of different ML docking methods on the PoseBusters test set}
    \label{fig:docking_baseline}
\end{figure*}

\begin{table*}
\caption{Performance on the posebusters and Astex diverse set for different traditional and ML models}
\vspace{8pt}
\centering
\begin{tabular}{c|cc}
    \toprule
       < 2.0 Å RMSD(\% ) & PoseBusters (N=428) & Astex (N=85) \\
        \midrule
        DeepDock & 17.8 & 34.12 \\ 
        DiffDock & 37.9 & 71.76 \\
        UMol & 45 & - \\
        Vina & 52.3 & 57.65 \\
        Uni-Mol Docking & 58.9 & 82.35 \\
        AlphaFold latest & 73.6 & - \\
        \midrule
        Uni-Mol Docking V2 & 77.6 & 95.29 \\
    \bottomrule
\end{tabular}
\label{tab:baseline_performance}
\end{table*}

\subsection{Increased Chemical Accuracy}

\begin{figure*}[t]
    \centering
    \includegraphics[width=6.5in]{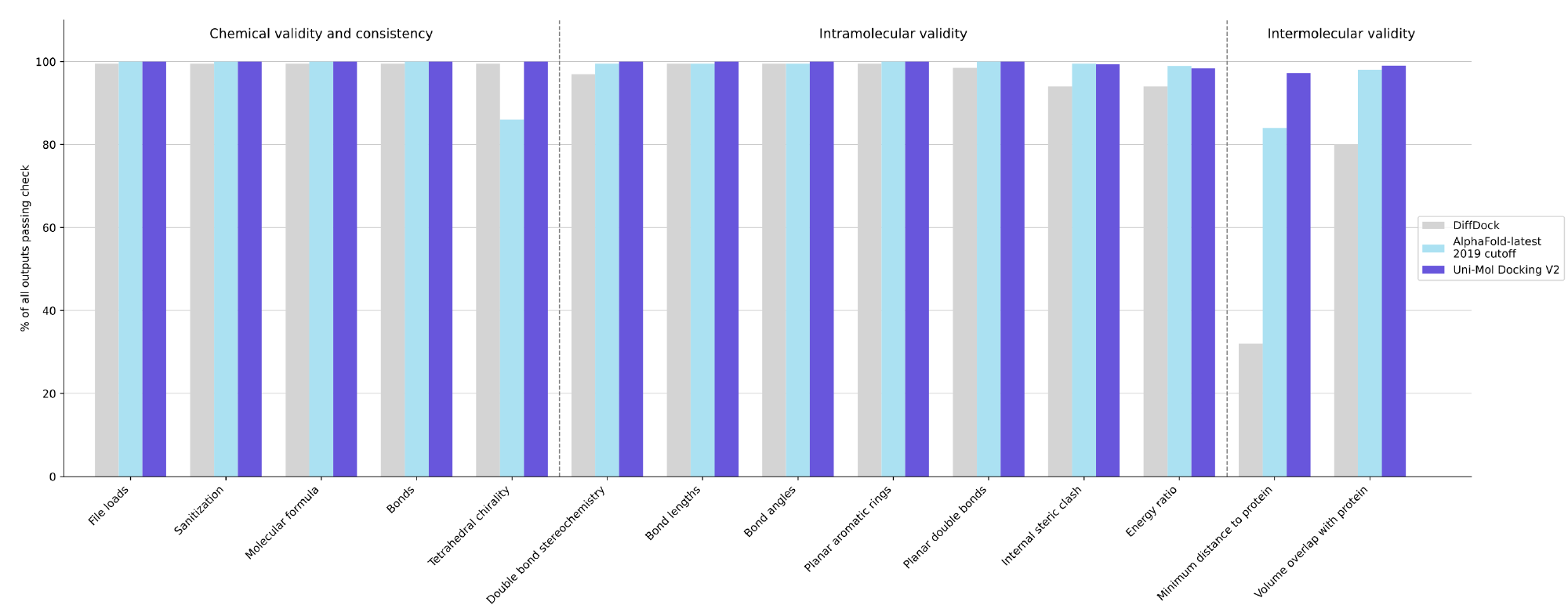}
    \caption{Comparison of plausibility checks for predictions by DiffDock, AlphaFold latest, and Uni-Mol Docking V2}
    \label{fig:plausibility_check}
\end{figure*}

\begin{figure*}
    \centering
    \includegraphics[width=6.5in]{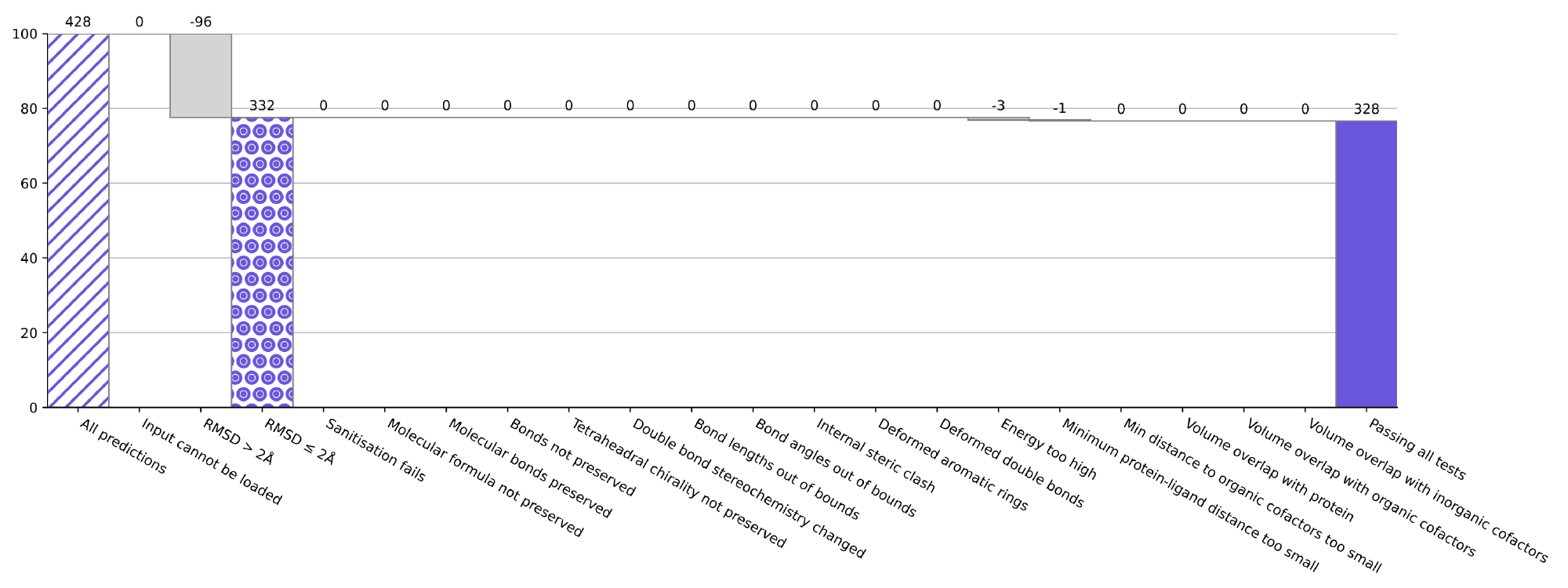}
    \caption{Waterfall plot showcasing cumulative impact of errors}
    \label{fig:plausibility_check_waterfall}
\end{figure*}

Moreover, we highlight Uni-Mol Docking V2 produces chemically accurate predictions, showcasing no chirality inversions nor steric clashes, unlike previous ML models. All issues presented in \cite{buttenschoen2023posebusters} have been resolved. 95+\% of predictions by Uni-Mol Docking V2 are chemically and physically plausible (as seen in figure 2 and 3).

We also report enhanced performance in high-quality predictions (RMSD <1.0 \AA~ and <1.5 \AA) and increased physical soundness when Uni-Mol Docking V2 is integrated with physics-based approaches like Uni-Dock. This setup enhances the industrial applications to rational drug design and virtual screening, with higher overall accuracy, reduced risk for overfitting,  bigger ratio of high quality predictions, and the integration of additional information of the binding site such as cofactors and crystalographic waters \cite{yang_lin_yuan_li_zou_zhou_zhang_zheng_2023}.

\subsection{Case Study}
We demonstrate the wide coverage of biochemical space in a wide variety of targets included in the PoseBusters test set with diverse biological functions and of interest in biotechnology, pharmaceutical and medical fields.

\begin{figure*}
    \centering
    \vspace{-0.75cm} % Adjust the value as needed
    \includegraphics[width=6.5in]{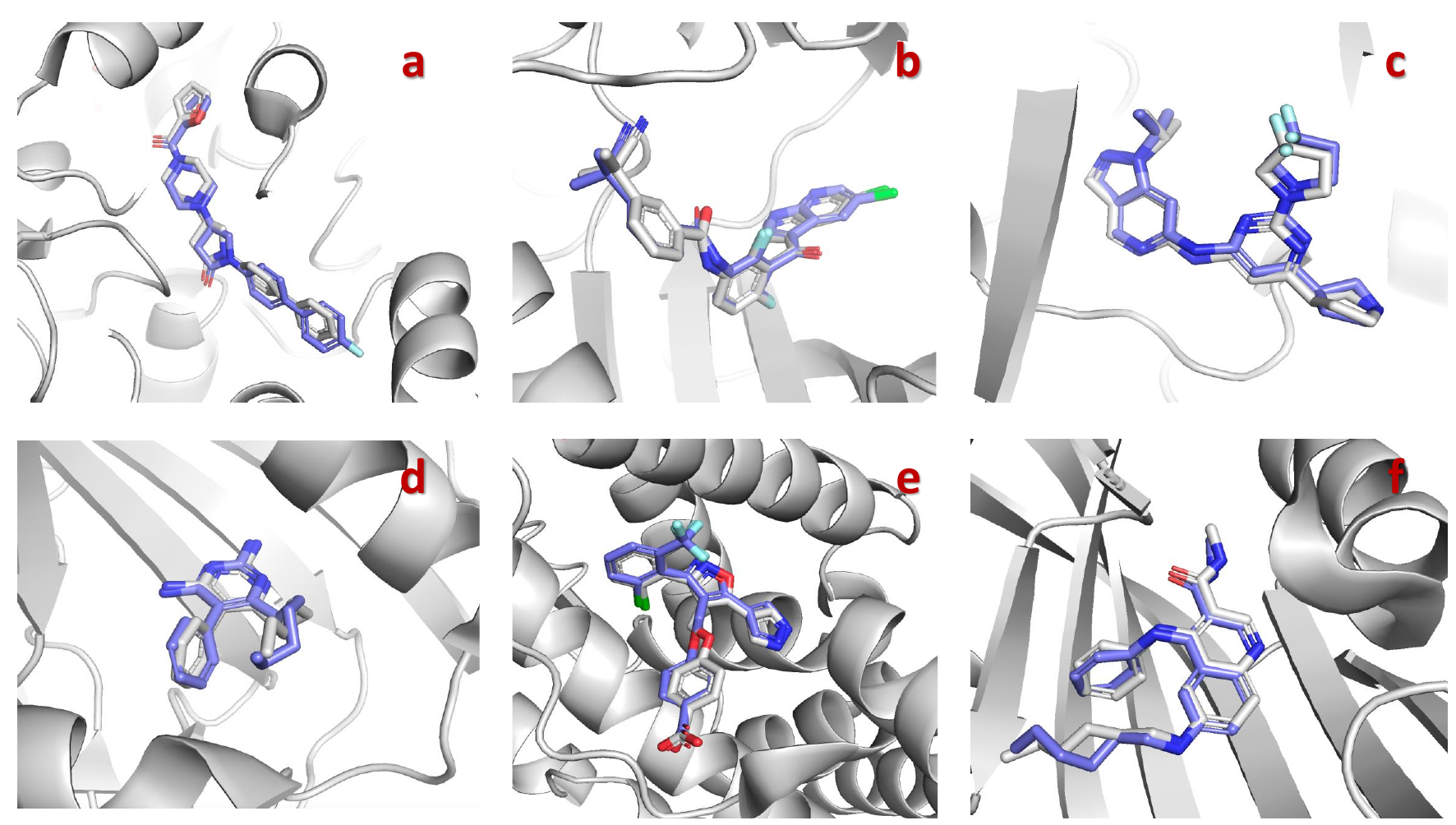}
    \caption{\textbf{a}: 7PRM, RMSD=1.11 \AA. Inhibition of Monoacylglycerol lipase (MAGL), a key enzyme
in the endocannabinoid system, has been proposed as an attractive approach for the treatment of
various diseases including neurodegeneration, psychiatric disorders, and cancer; \textbf{b}: 8C7Y,
RMSD=0.38 \AA. BRAF inhibitors have revolutionized treatment of some cancers such as melanoma,
although some undesired effects have been seen, such as the paradoxical hyperactivation of MAPK
caused by the ligand-induced dimerization of 1st gen BRAF inhibitors. \textbf{c}: 7R9N. RMSD=0.63 \AA.
Hematopoietic progenitor kinase 1 (HPK1) is implicated as a negative regulator of T-cell
receptor-induced T-cell activation. Inhibition of HPK1 has been shown to increase T-cell antitumor
response. \textbf{d}: 7XI7, RMSD=1.25 \AA. Novel inhibitors for human dihydrofolate reductase could
expand therapeutic options against the parasitic toxoplasmosis infections. \textbf{e}: 7NP6, RMSD=0.54 \AA.
Inhibition of the nuclear receptor retinoic-acid-receptor-related orphan receptor $\gamma \text{t}$ (ROR$\gamma \text{t}$) is a
promising strategy in the treatment of autoimmune diseases. \textbf{f}: 7N03, RMSD=1.03 \AA. MTH1 is a
DNA damage control enzyme and potentially synthetic lethal target. Its inhibition could open new
avenues in oncologic targeted therapy.}
    \label{fig:bindpose_case}
\end{figure*}

%% file: sections/conclusion.tex
\section{Conclusion}
\vspace{-0.1cm}
The latest version of UniMol Docking establishes a new state of the art on the PoseBusters benchmark, surpassing its predecessor, which was the top-performing open-source model to the best of our knowledge as of November 22, 2023. 
Our findings signify a new frontier in the application of AI for Science in Molecular Docking. This is achieved through a comprehensive approach that addresses the ligand docking problem and rectifies implausible outcomes previously generated by machine learning models.
The combination of deep learning methods and physics-based methods improves performance and allows for the incorporation of extra information, making the pipeline more suitable for industrial applications in Virtual Screening and Drug Design. 

%% file: sections/code_data.tex
\vspace{-0.2cm} % Adjust the value as needed
\section{Code and Data Availability}
\vspace{-0.15cm} % Adjust the value as needed
All Code and data is availabe for public, also we provide Uni-Mol Docking V2 service for non-commercial usage:
\vspace{-0.15cm} % Adjust the value as needed
\begin{enumerate}
    \item The previous version of Uni-Mol Docking, achieving 62\% <2 \AA~RMSD on the posebusters set, along with the Posebusters and Astex datasets can be accessed in GitHub via \hyperlink{explicit reproduction instructions}{https://github.com/dptech-corp/Uni-Mol/pull/178}.
    \item Uni-Mol Docking V2, achieving 77.62\% <2 \AA~RMSD on the posebusters set, along with model weight, can be accessed in GitHub via \hyperlink{Uni-Mol Docking fold}{https://github.com/dptech-corp/Uni-Mol/tree/main/unimol\_docking\_v2}.
    \item Uni-Mol Docking V2 service is available as a preliminary demo via \hyperlink{Bohrium Apps}{https://bohrium.dp.tech/apps/unimoldockingv2}.
    \item Prepared protein-ligand complex from MOAD is available via \hyperlink{Zendo}{https://zenodo.org/records/11191555}.
\end{enumerate}

%% file: sections/contribution.tex
\section*{Contributions and Acknowledgments}

\subsection{Contributors(in alphabetical orders)}

\textbf{Eric Alcaide}: Conceptualization, Methodology, Software, Validation, Writing.

\textbf{Zhifeng Gao}: Supervision, Conceptualization, Methodology, Writing.

\textbf{Guolin Ke}: Supervision, Project administration, Conceptualization, Methodology.

\textbf{Yaqi Li}: Data Curation, Investigation, Validation.

\textbf{Linfeng Zhang}: Supervision, Funding acquisition.

\textbf{Hang Zheng}: Software, Methodology, Data Curation, Validation.

\textbf{Gengmo Zhou}: Conceptualization, Methodology, Software, Validation, Writing.

\subsection{Acknowledgments}
Our work is made possible by the dedication and
efforts of numerous teams at DP Technology and AI for Science Institute. We would like to acknowledge the support from MLOps Team, PR Team, Drug Discovery Team etc.